\documentclass[%
 reprint,
 amsmath,amssymb,
 aps,
]{revtex4-2}

\usepackage{graphicx}
\usepackage{dcolumn}
\usepackage{bm}

\begin{document}

\preprint{APS/123-QED}

\title{A Monte Carlo approach to model COVID-19 deaths and infections using Gompertz functions}

\author{Tulio Rodrigues}
\email{tulio@if.usp.br}
\author{Otaviano Helene}
\affiliation{
 Experimental Physics Department\\
Physics Institute - University of São Paulo - Brazil}

\date{\today}

\begin{abstract}
This study provides a phenomenological method to describe the exponential growth, saturation and decay of coronavirus disease 2019 (COVID-19) deaths and infections via a Monte Carlo approach. The calculations connect Gompertz-type trial distributions of infected people per day with the distribution of deaths adopting two gamma distributions to account for the elapsed time that encompass the incubation and symptom onset to death periods. The analyses include death's data from United States of America (USA), Brazil, Mexico, United Kingdom (UK), India and Russia, which comprise the four countries with the highest number of deaths and the four countries with the highest number of confirmed cases, as of Aug 07, 2020, according to the World Health Organization webpage: \url{https://covid19.who.int/table}. The Gompertz functions were fitted to the data of weekly averaged confirmed deaths per day by mapping the $\chi^2$ values. The uncertainties, variances and covariances of the model parameters were calculated by propagation, taking into account the standard deviations of the data for each epidemiological week. The fitted functions for the average deaths per day for USA and India have an upward trend, with the former having a higher growth rate and quite huge uncertainties. For Mexico, UK and Russia, the fits are consistent with a slope down pattern. For Brazil we found a subtle trend down, but with significant uncertainties. The USA, UK and India data shown a first peak with a higher growth rate when compared to the second one (typically 2.7 to 3.7 times higher), demonstrating the benefits of non-pharmaceutical interventions of sanitary measures and social distance flattening the curves of the pandemic. For the case of USA, however, a third peak seems quite plausible, most likely related with the recent relaxation policies. Brazil's data are satisfactorily described by two highly overlapped Gompertz functions with similar growth rates, suggesting a two-steps process for the pandemic spreading. For the case of Mexico and Russia single peaks with smoother slopes fitted the data satisfactorily. The 95\% confidence intervals for the total number of deaths $(\times10^3)$ predicted by the model for Aug 31, 2020 are 160 to 220, 110 to 130, 59 to 62, 46.6 to 47.3, 54 to 63 and 16.0 to 16.7 for USA, Brazil, Mexico, UK, India and Russia, respectively. Our estimates for the prevalences of infections are in reasonable agreement with some preliminary reports from serological studies carried out in USA and Brazil. The prevalences and 95\% confidence intervals for Aug 1, 2020 were estimated to be 8.3(5.7-10.9)\%, 7.9(6.6-9.2)\%, 6.7(6.0-7.5)\%, 10.5(8.4-12.6)\%, 0.6(0.5-0.7)\% and 1.6(1.5-1.8)\% for USA, Brazil, Mexico, UK, India and Russia, respectively. The method represents an effective framework to estimate the line-shape of the infection curves and the uncertainties of the relevant parameters based on the actual data, in contrast with more complex epidemiological models that require a comprehensive knowledge of several parameters.  

\end{abstract}

\maketitle


The outbreak of the new coronavirus disease 2019 (COVID-19) brought a challenging scenario worldwide \cite{wu2020new,bedford2019new,lancet2020emerging}, urging timely and effective responses from the authorities regarding the availability of intensive care units \cite{phua2020intensive,christen2020report}, as well as the implementation of non-pharmaceutical interventions of social distance and protective sanitary measures
\cite{ferguson2020impact,kraemer2020effect,wang2020response}. Epidemiological models \cite{kucharski2020early,overton2020using,giordano2020modelling,prem2020effect, gaeta2020simple, scabini2020social, leung2020first,candido2020evolution,fang2020transmission,massad2020two} and other statistical approaches \cite{schuttler2020covid,pedrosa2020dynamics,ul2020understanding} have been very useful to guide actions to manage this crisis and to shed light on how to safely and gradually resume economics and social activities \cite{xu2020analysis}. On the other hand, quantitative analyses are strongly susceptible to several uncertainties, such as under-reporting of confirmed cases and deaths \cite{paixao2020estimation,krantz2020level, jagodnik2020correcting}, lack of massive tests in some countries \cite{cohen2020countries}, changes in policies and methods for reporting confirmed cases and deaths as time evolves during the pandemic growth, very distinct socio-economic patterns and health facilities capabilities among different countries and also among different focuses of the disease within the same country. Such complex and puzzling scenario directly influences the forecast capabilities of the calculations, supporting the need of a multidisciplinary cooperation of the scientific community and the development of mathematical models to provide plausible estimates for the uncertainties of the relevant parameters \cite{capaldi2012parameter,gaffey2018application,kuniya2020prediction,smirnova2019stable}. Therefore, the present analysis provides an effective phenomenological method to estimate the magnitude and the relevant uncertainties of some important quantities as the pandemic evolves, such as: (1) the peak day(s), growth rate(s) and total number of infected people for the reconstructed infection curves, (2) forecast distribution of deaths per day, and (3) forecast total number of deaths until Aug 31, 2020. It is worth mentioning, however, that the predictions of the model strongly depend on the prevailing conditions already in place for the specific country which data were analyzed, as any substantial change in governmental policies, either in the direction of loosening or tightening social distance, will generate different dynamics for the spread of the virus.\\
The best-fit parameters are generally strongly correlated, but its uncertainties reflect the dispersion of the data at each epidemiological week, making the present analysis a suitable quantitative method to describe the pandemic line-shape behavior with few parameters. Moreover, such approach could also be useful for the identification of upward trends and its correlations with relaxation policies and procedures as global social and economics activities are gradually resumed \cite{peto2020covid}.\\
The calculations assume that the total number of infected people increase according with a Gompertz-type function, which is a sigmoid curve with a lower growth rate at the beginning and at the end, such that:
\begin{equation}
I(t)=Ne^{-e^{-\lambda (t-t_{0})}},
\end{equation}%
where $N$ represents the asymptotic number of infected people for $t\rightarrow\infty$, $\lambda$ is the growth rate and $t_{0} $ the peak time of the derivative of $I(t)$. In such model, the number of infected people per time period can be written as:
\begin{equation}
G(t) = \frac{d}{dt}I(t) =N\lambda e^{-e^{-\lambda(t-t_{0})}}e^{-\lambda(t-t_{0})}.
\end{equation}

For the specific case of COVID-19, the data of confirmed cases depend very strongly on testing and reporting policies for the related country. These policies may also vary along time, distorting the shape of the distributions. This complicated scenario disfavor the use of confirmed cases as a reliable source of information to describe the pandemic dynamics, which is crucial to guide government actions and decisions. In order to overcome these difficulties, we have adopted the data of deaths per day, as they should be more consistent with the actual spreading mechanism of the virus.
The connection between the trial function of the number of infected people per day and the number of deaths per day takes into account the probability distribution function for the elapsed time between the infection and the death, which can be satisfactorily described by the sum of two time periods, namely, the incubation period $t_{inc}$, and the symptom onset to death period $t_{s-d}$. Both periods are generated in the Monte Carlo algorithm assuming that they are independent and gamma distributed with an average of 5.1 and 17.8 days and coefficient of variation of 0.86 and 0.45, respectively, as proposed elsewhere \cite{flaxman2020report} based on previous studies of Wuhan data \cite{ferguson2020impact,verity2020estimates}\footnote{The average elapsed time from symptom onset to death was corrected to 17.8 days in \protect\cite{verity2020estimates}, instead of their previous estimates of 18.8 days used by \protect\cite{flaxman2020report}}. So, the time of the death $t_{d}$ can be written as: $t_{d}=t_{inf}+t_{inc}+t_{s-d}$, with $t_{inf}$ representing the time of infection.\\
The analysis of the deaths data \footnote{The deaths data included in the analysis were taken from the European Centre for Disease Prevention and Control webpage: https://www.ecdc.europa.eu/en/publications-data/download-todays-data-geographic-distribution-covid-19-cases-worldwide} were done considering the weekly averaged deaths per day and its corresponding standard deviation of the average for the respective epidemiological week, counted retrospectively from the end date of Aug 7, 2020.
The time counting considered the day of the first confirmed case for each country and the first epidemiological weeks were chosen in such a way that all days of the week had at least one death, meaning that the present calculations do consider the early stages of the pandemic, including from 19 up to 22 weeks depending on the country (see Table~\ref{tab:table1}). The mean day of each epidemiological week was chosen with a time bin of $\pm$ 3.5 days, and the mean of each day was taken at the half of the day.
The trial line-shapes of the number of infected people per day were generated considering a single Gompertz function for the case of Mexico and Russia, a sum of two Gompertz functions for Brazil, United Kingdom, and India and three Gompertz functions for USA. Such approach allows the inclusion of two or more superimposed dynamics of the disease as one would expect considering the changes in policies as time evolves (which might remarkably reduce the growth rates) and also the cluster structure expected for large countries with multiple focuses of the disease (not observed for Mexico and Russia so far). Such method would also allow the inclusion of other peaks as the countries start their processes of re-opening social and economics activities, which seems to be the case for USA.

\begin{figure*}
\includegraphics[scale=0.65]{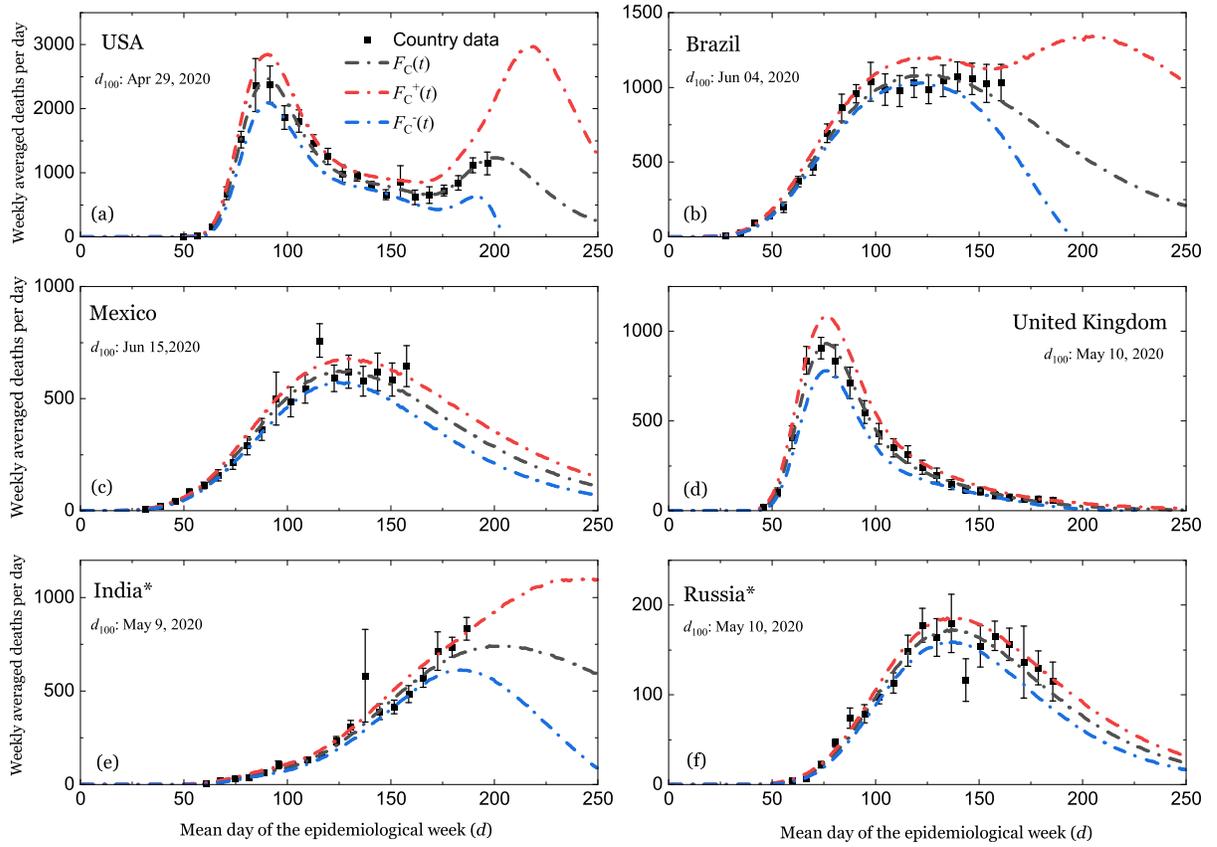}
\caption{\label{fig:Fig1}Weekly averaged deaths per day for USA (a), Brazil (b), Mexico (c), United Kingdom (d), India (e) and Russia (f) (data points) and the respective model estimations given by $F_{C}(t) $ (Gray dotted-dashed lines) and its 95\% CI upper and lower limits (red and blue dashed dotted lines, respectively). Also shown for clarity the hundredths calendar day ($ d_{100}$) since the first confirmed case for each country. (*) For the case of India and Russia a 5\% error was added to the death's data to achieve a successful fitting.}
\end{figure*}

\begin{figure*}
\includegraphics[scale=0.5]{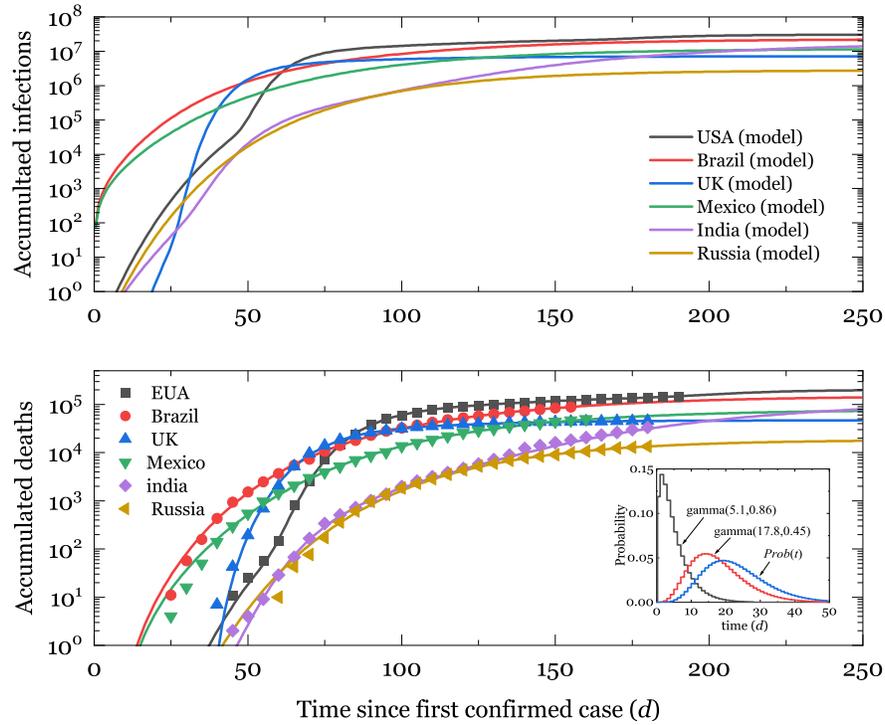}
\caption{\label{fig:Fig2}Upper panel: Model predictions for the accumulated number of infections for all six countries (solid lines). Lower Panel: Model predictions for the accumulated deaths (solid lines) versus the available data (data-points), which are presented within 5 days' time intervals for clarity. The insert of the lower panel shows the Monte Carlo generated distribution \textit{Prob}(\textit{t}) (blue histogram) and the gamma distributions for the incubation (black histogram) and symptom onset to death (red histogram) periods.}
\end{figure*}

The fitting procedure was performed by mapping the $\chi^2$, defined as:
\begin{equation}
\chi^2=\sum_{iw=1}^{n}\frac{[ \tilde{F}_{iw} - y_{iw}] ^{2}}{\sigma y_{iw}^{2}},
\end{equation}
where $\tilde{F}_{iw}$ is the trial function for deaths calculated at time $t_{iw}$ (the mean day of the corresponding week), $y_{iw}$ the weekly averaged deaths per day and $\sigma y_{iw}$ its standard deviation. In that sense, the average values with higher standard deviations had lower weights in the fitting procedure.  
The best fit parameters of the Gompertz functions ($N, \lambda$ and $t_{0}$) where obtained by sorting 5000 random sets of parameters around plausible guessing values and calculating the respective trial function and $\chi ^2$ for each candidate, assuming a total of 10 million infections using a time bin of one day. The trial functions $\tilde{F}_{iw}$ were calculated for each set of parameters by the connection between the 10 million infection events with the respective death events using the probability distribution function of $t_{inc}+t_{s-d}$, herein denoted $Prob(\Delta t) = Prob(t_{d}-t_{inf})$ (see the insert of Fig.\ref{fig:Fig2}). This procedure was done several times with progressively narrow bins for each parameter's increments until the respective $\chi ^2$ converged to the minimal value (the convergence criteria required that the $\chi^2$ obtained for 5000 random sets of parameters is lower than the $\chi^2$ obtained for 4000 runs and their difference is lower than 0.05 units). The calculation of the total number of infected people $N$ (three Gompertz functions for USA; two for Brazil, UK and India and one for Mexico and Russia) was performed assuming an infection-fatality-ratio (\textit{ifr}) of 0.66\% \cite{verity2020estimates}.
A least square method \cite{vanin2007covariance} was applied for the calculation of the uncertainties of the best fit parameters, which covariance matrix can be written as:
\begin{equation}
Vb=(\tilde{F}^{'\top} V^{-1} \tilde{F}^{'}) ^{-1},
\end{equation}
where $\tilde{F}=\tilde{F}_{iw,j}^{'}$ stands for the partial derivative of $\tilde{F}_{iw}$ at any $t_{iw}$ in respect to the $P_{j}$ parameter of each Gompertz function ($N, \lambda$ and $t_{0}$). The variance matrix of the death's data is diagonal, such that: $V_{iw,iw} = (\sigma y_{iw} ^{2}) ^{-1} $. Given the lower number of deaths per day for India and Russia and the huge variation of the data in each week, we have included an additional uncertainty of 5\% ($\sigma y_{iw} \rightarrow \sigma y_{iw}+0.05y_{iw}$) in order to achieve a successful fitting. For the calculation of $\tilde{F}^{'}_{iw,j}$ we have used the resulting convolution between the reconstructed curve of infected people $G(t)$ and the probability density function $Prob(\Delta t)$, such that:
\begin{equation}
\tilde{F}^{'}_{iw,j} = \frac{d}{dP_{j}}[F_{C}(t_{iw},N_{k}, \lambda_{k}, t_{0k})] \text{, with}
\end{equation}
\begin{equation}
F_{C}(t) = ifr\cdot\int_{0}^{t} \sum_{k=1}^{k_{max}}[G(\tau,N_{k}, \lambda_{k}, t_{0k})Prob(t-\tau)d\tau],
\end{equation}
where $k_{max}=1$ for Mexico and Russia, 2 for Brazil, UK and India and 3 for USA.
The propagation of the uncertainties of the best fit parameters to the reconstructed infection curve took into account the full co-variance matrix $Vb$, as similarly described in \citep{helene2016useful}, with the vector $G^{'}_{m}$ being defined as:
\begin{equation}
G^{'}_{m} = \begin{pmatrix}
G^{'}_{m,1}\\
\vdots\\
G^{'}_{m,j_{max}}
\end{pmatrix},
\end{equation}
where $ G^{'}_{m,1}, \texttt{...} G^{'}_{m,j_{max}}$ are the partial derivatives of the infection curve $ G (t_{m},N_{k}, \lambda_{k}, t_{0k})$ in respect to the parameter $P_{1}, \texttt{...} P_{j_{max}}$ calculated at each day $t_{m}$ ($ j_{max} = 3$ for Mexico and Russia, 6 for Brazil, UK and India and 9 for USA). Consequently, the uncertainty of the infection curve at each point can be written as:
\begin{equation}
\sigma G_{m} =\sqrt{G_{m}^{'\top}VbG^{'}_{m}}.
\end{equation}
The uncertainties in the convoluted functions can be calculated as:
\begin{equation}
F_{C}^{\pm}(t) = F_{C}(t) \pm ifr\cdot\int_{0}^{t} [\sigma G(\tau)Prob(t-\tau)d\tau],
\end{equation}
where $\sigma G(\tau)$ is obtained by the interpolation of $\sigma G_{m}$. 
Figure~\ref{fig:Fig1} shows the weekly averaged deaths per day distributions for all six countries (data points) and its respective convoluted functions $ F_{c}(t) $ (dashed-dotted gray lines). The upper and lower estimates $F_{C}^{\pm}(t)$ [95\% Confidence Intervals(CI)] are presented by the red and blue dashed-dotted lines, respectively.
The total number of deaths and its 95\% CI at any given time $t_{f} $ can be written as:
\begin{equation}
N_{d}(t_{f})= N_{d}(t_{i})+ \int_{t_{i}+1}^{t_{f}}F_{C}(t')dt' \text{, and}
\end{equation}
\begin{equation}
N_{d}^{\pm}(t_{f})= N_{d}(t_{i})+ \int_{t_{i}+1}^{t_{f}}F_{C}^{\pm}(t')dt',
\end{equation}
where $N_{d}(t_{i})$ corresponds to the actual data of accumulated deaths until the day $(t_{i})$ for each country.\\
Table~\ref{tab:table1} summarizes all the results, including the model predictions for the total number of deaths and its 95\% CI for Aug 31, 2020. The countries with well defined first peaks (USA and UK) present the highest initial growth rates ($ 0.109\pm0.012 d^{-1}$ and $0.100\pm0.017 d^{-1}$) and uncertainties in the peak days of 1.0 and 2.2 days, respectively. The second peaks in both cases have much lower growth rates, demonstrating the flattening of the infection curve due to non-pharmacological interventions of social distance and sanitary measures. Besides UK, which clearly shows a well controlled scenario, Mexico and Russia also have a trend down with modest uncertainties, which is a consequence of the fitting being successfully performed with a single Gompertz function. For the case of India, the first Gompertz function has a quite small contribution ($\sim$2\%) in the total number of infections and the second curve dominates the distribution of deaths per day; resulting in modest uncertainties. For the case of Brazil, there are few data points to properly constraint the second Compertz function, which has a similar growth rate compared with the first one. The large overlap between the two functions leads to high correlation coefficients between $ N_{1} $ and $ \lambda _{1} $(-0.989), $ N_{2} $ and $ \lambda _{2} $(-0.972),  $ \lambda _{1} $ and  $\lambda _{2} $(-0.841) and $ N_{1} $ and $ N_{2} $ (-0.973), influencing for the large uncertainties. A similar situation also play a role for the huge uncertainties found in the parameters of the third Gompertz function for USA, which is weakly constrained with few data points. The death's peak days have an average shift of 22.9 days from the corresponding infection's peaks, which is the average of \textit{Prob}(\textit{t}) (the sum of the averages of the two gamma functions).
The estimates for the prevalences of infections at the beginning of each month are also shown in Table ~\ref{tab:table1} for all six countries. For the case of USA, our estimated prevalence for April 3-4, 2020 [2.6\% (95\% CI, 2.1 to 3.0\%)] is in good agreement with a preliminary serological study carried out in Santa Clara County \cite{bendavid2020covid} 2.8\% (95\% CI, 1.3-4.7\%). For the case of Brazil, we found a prevalence of 2.6\% (95\% CI 2.3-2.9\%) within May 15-22, 2020, which is higher than the overall prevalence found in a survey that included 90 cities of Brazil 1.4\% (95\% CI, 1.3-1.6\%)\cite{hallal2020remarkable}. On the other hand, our result for May 14, 2020 [(2.3\%(95\% CI, 2.1-2.6\%)] is in good agreement with the figures reported in a preliminary research performed in the Brazilian State of Espirito Santo \cite{gomes2020population} 2.1\% (95\% CI, 1.67-2.52\%). Obviously that these comparisons should be done with parsimony, given the fact that our results refer to an overall estimate for each country and are strictly related with an infection-fatality-ratio of 0.66\% \cite{verity2020estimates}.  
       
The upper panel of Figure~\ref{fig:Fig2} shows the model predictions for the accumulated number of infections (solid lines) and the lower panel shows the corresponding model estimates for the accumulated deaths, in comparison with the available data (data points) at each 5 day time interval. Once again it is verified a nice agreement between the data and the model, with some discrepancies found in the early stages of the pandemic (accumulated number of deaths $\lesssim$ 100). The Monte Carlo generated probability distribution \textit{Prob}(\textit{t}), and the two gamma distributions (incubation and symptom onset to deaths periods) are presented in the insert of the lower panel of Figure~\ref{fig:Fig2}.\\

In conclusion, we have presented a few-parameter model to describe the dynamics of a pandemic in terms of Gompertz functions using Monte Carlo techniques to determine the best fit parameters and a least square method - weighted by the dispersion of the death's data in each epidemiological week - to estimate the relevant uncertainties. 

\bibliography{manuscript_covid19_rodrigues_helene}

\begin{thebibliography}{40}%
\makeatletter
\providecommand \@ifxundefined [1]{%
 \@ifx{#1\undefined}
}%
\providecommand \@ifnum [1]{%
 \ifnum #1\expandafter \@firstoftwo
 \else \expandafter \@secondoftwo
 \fi
}%
\providecommand \@ifx [1]{%
 \ifx #1\expandafter \@firstoftwo
 \else \expandafter \@secondoftwo
 \fi
}%
\providecommand \natexlab [1]{#1}%
\providecommand \enquote  [1]{``#1''}%
\providecommand \bibnamefont  [1]{#1}%
\providecommand \bibfnamefont [1]{#1}%
\providecommand \citenamefont [1]{#1}%
\providecommand \href@noop [0]{\@secondoftwo}%
\providecommand \href [0]{\begingroup \@sanitize@url \@href}%
\providecommand \@href[1]{\@@startlink{#1}\@@href}%
\providecommand \@@href[1]{\endgroup#1\@@endlink}%
\providecommand \@sanitize@url [0]{\catcode `\\12\catcode `\$12\catcode
  `\&12\catcode `\#12\catcode `\^12\catcode `\_12\catcode `\%12\relax}%
\providecommand \@@startlink[1]{}%
\providecommand \@@endlink[0]{}%
\providecommand \url  [0]{\begingroup\@sanitize@url \@url }%
\providecommand \@url [1]{\endgroup\@href {#1}{\urlprefix }}%
\providecommand \urlprefix  [0]{URL }%
\providecommand \Eprint [0]{\href }%
\providecommand \doibase [0]{https://doi.org/}%
\providecommand \selectlanguage [0]{\@gobble}%
\providecommand \bibinfo  [0]{\@secondoftwo}%
\providecommand \bibfield  [0]{\@secondoftwo}%
\providecommand \translation [1]{[#1]}%
\providecommand \BibitemOpen [0]{}%
\providecommand \bibitemStop [0]{}%
\providecommand \bibitemNoStop [0]{.\EOS\space}%
\providecommand \EOS [0]{\spacefactor3000\relax}%
\providecommand \BibitemShut  [1]{\csname bibitem#1\endcsname}%
\let\auto@bib@innerbib\@empty
\bibitem [{\citenamefont {Wu}\ \emph {et~al.}(2020)\citenamefont {Wu},
  \citenamefont {Zhao}, \citenamefont {Yu}, \citenamefont {Chen}, \citenamefont
  {Wang}, \citenamefont {Song}, \citenamefont {Hu}, \citenamefont {Tao},
  \citenamefont {Tian}, \citenamefont {Pei} \emph {et~al.}}]{wu2020new}%
  \BibitemOpen
  \bibfield  {author} {\bibinfo {author} {\bibfnamefont {F.}~\bibnamefont
  {Wu}}, \bibinfo {author} {\bibfnamefont {S.}~\bibnamefont {Zhao}}, \bibinfo
  {author} {\bibfnamefont {B.}~\bibnamefont {Yu}}, \bibinfo {author}
  {\bibfnamefont {Y.-M.}\ \bibnamefont {Chen}}, \bibinfo {author}
  {\bibfnamefont {W.}~\bibnamefont {Wang}}, \bibinfo {author} {\bibfnamefont
  {Z.-G.}\ \bibnamefont {Song}}, \bibinfo {author} {\bibfnamefont
  {Y.}~\bibnamefont {Hu}}, \bibinfo {author} {\bibfnamefont {Z.-W.}\
  \bibnamefont {Tao}}, \bibinfo {author} {\bibfnamefont {J.-H.}\ \bibnamefont
  {Tian}}, \bibinfo {author} {\bibfnamefont {Y.-Y.}\ \bibnamefont {Pei}}, \emph
  {et~al.},\ }\bibfield  {title} {\bibinfo {title} {A new coronavirus
  associated with human respiratory disease in {C}hina},\ }\href@noop {}
  {\bibfield  {journal} {\bibinfo  {journal} {Nature}\ }\textbf {\bibinfo
  {volume} {579}},\ \bibinfo {pages} {265} (\bibinfo {year}
  {2020})}\BibitemShut {NoStop}%
\bibitem [{\citenamefont {Bedford}\ \emph {et~al.}(2019)\citenamefont
  {Bedford}, \citenamefont {Farrar}, \citenamefont {Ihekweazu}, \citenamefont
  {Kang}, \citenamefont {Koopmans},\ and\ \citenamefont
  {Nkengasong}}]{bedford2019new}%
  \BibitemOpen
  \bibfield  {author} {\bibinfo {author} {\bibfnamefont {J.}~\bibnamefont
  {Bedford}}, \bibinfo {author} {\bibfnamefont {J.}~\bibnamefont {Farrar}},
  \bibinfo {author} {\bibfnamefont {C.}~\bibnamefont {Ihekweazu}}, \bibinfo
  {author} {\bibfnamefont {G.}~\bibnamefont {Kang}}, \bibinfo {author}
  {\bibfnamefont {M.}~\bibnamefont {Koopmans}},\ and\ \bibinfo {author}
  {\bibfnamefont {J.}~\bibnamefont {Nkengasong}},\ }\bibfield  {title}
  {\bibinfo {title} {A new twenty-first century science for effective epidemic
  response},\ }\href@noop {} {\bibfield  {journal} {\bibinfo  {journal}
  {Nature}\ }\textbf {\bibinfo {volume} {575}},\ \bibinfo {pages} {130}
  (\bibinfo {year} {2019})}\BibitemShut {NoStop}%
\bibitem [{\citenamefont {Lancet}(2020)}]{lancet2020emerging}%
  \BibitemOpen
  \bibfield  {author} {\bibinfo {author} {\bibfnamefont {T.}~\bibnamefont
  {Lancet}},\ }\bibfield  {title} {\bibinfo {title} {Emerging understandings of
  2019-ncov},\ }\href@noop {} {\bibfield  {journal} {\bibinfo  {journal}
  {Lancet (London, England)}\ }\textbf {\bibinfo {volume} {395}},\ \bibinfo
  {pages} {311} (\bibinfo {year} {2020})}\BibitemShut {NoStop}%
\bibitem [{\citenamefont {Phua}\ \emph {et~al.}(2020)\citenamefont {Phua},
  \citenamefont {Weng}, \citenamefont {Ling}, \citenamefont {Egi},
  \citenamefont {Lim}, \citenamefont {Divatia}, \citenamefont {Shrestha},
  \citenamefont {Arabi}, \citenamefont {Ng}, \citenamefont {Gomersall} \emph
  {et~al.}}]{phua2020intensive}%
  \BibitemOpen
  \bibfield  {author} {\bibinfo {author} {\bibfnamefont {J.}~\bibnamefont
  {Phua}}, \bibinfo {author} {\bibfnamefont {L.}~\bibnamefont {Weng}}, \bibinfo
  {author} {\bibfnamefont {L.}~\bibnamefont {Ling}}, \bibinfo {author}
  {\bibfnamefont {M.}~\bibnamefont {Egi}}, \bibinfo {author} {\bibfnamefont
  {C.-M.}\ \bibnamefont {Lim}}, \bibinfo {author} {\bibfnamefont {J.~V.}\
  \bibnamefont {Divatia}}, \bibinfo {author} {\bibfnamefont {B.~R.}\
  \bibnamefont {Shrestha}}, \bibinfo {author} {\bibfnamefont {Y.~M.}\
  \bibnamefont {Arabi}}, \bibinfo {author} {\bibfnamefont {J.}~\bibnamefont
  {Ng}}, \bibinfo {author} {\bibfnamefont {C.~D.}\ \bibnamefont {Gomersall}},
  \emph {et~al.},\ }\bibfield  {title} {\bibinfo {title} {Intensive care
  management of coronavirus disease 2019 ({COVID-19}): challenges and
  recommendations},\ }\href@noop {} {\bibfield  {journal} {\bibinfo  {journal}
  {The Lancet Respiratory Medicine}\ } (\bibinfo {year} {2020})}\BibitemShut
  {NoStop}%
\bibitem [{\citenamefont {Christen}\ \emph {et~al.}(2020)\citenamefont
  {Christen}, \citenamefont {D'Aeth}, \citenamefont {Lochen}, \citenamefont
  {McCabe}, \citenamefont {Rizmie}, \citenamefont {Schmit}, \citenamefont
  {Nayagam}, \citenamefont {Miraldo}, \citenamefont {White}, \citenamefont
  {Aylin} \emph {et~al.}}]{christen2020report}%
  \BibitemOpen
  \bibfield  {author} {\bibinfo {author} {\bibfnamefont {P.}~\bibnamefont
  {Christen}}, \bibinfo {author} {\bibfnamefont {J.}~\bibnamefont {D'Aeth}},
  \bibinfo {author} {\bibfnamefont {A.}~\bibnamefont {Lochen}}, \bibinfo
  {author} {\bibfnamefont {R.}~\bibnamefont {McCabe}}, \bibinfo {author}
  {\bibfnamefont {D.}~\bibnamefont {Rizmie}}, \bibinfo {author} {\bibfnamefont
  {N.}~\bibnamefont {Schmit}}, \bibinfo {author} {\bibfnamefont
  {A.}~\bibnamefont {Nayagam}}, \bibinfo {author} {\bibfnamefont
  {M.}~\bibnamefont {Miraldo}}, \bibinfo {author} {\bibfnamefont
  {P.}~\bibnamefont {White}}, \bibinfo {author} {\bibfnamefont
  {P.}~\bibnamefont {Aylin}}, \emph {et~al.},\ }\href@noop {} {\emph {\bibinfo
  {title} {Report 15: Strengthening hospital capacity for the {COVID-19}
  pandemic}}},\ \bibinfo {type} {Tech. Rep.}\ (\bibinfo {year}
  {2020})\BibitemShut {NoStop}%
\bibitem [{\citenamefont {Ferguson}\ \emph {et~al.}(2020)\citenamefont
  {Ferguson}, \citenamefont {Laydon}, \citenamefont {Nedjati-Gilani},
  \citenamefont {Imai}, \citenamefont {Ainslie}, \citenamefont {Baguelin},
  \citenamefont {Bhatia}, \citenamefont {Boonyasiri}, \citenamefont
  {Cucunub{\'a}}, \citenamefont {Cuomo-Dannenburg} \emph
  {et~al.}}]{ferguson2020impact}%
  \BibitemOpen
  \bibfield  {author} {\bibinfo {author} {\bibfnamefont {N.~M.}\ \bibnamefont
  {Ferguson}}, \bibinfo {author} {\bibfnamefont {D.}~\bibnamefont {Laydon}},
  \bibinfo {author} {\bibfnamefont {G.}~\bibnamefont {Nedjati-Gilani}},
  \bibinfo {author} {\bibfnamefont {N.}~\bibnamefont {Imai}}, \bibinfo {author}
  {\bibfnamefont {K.}~\bibnamefont {Ainslie}}, \bibinfo {author} {\bibfnamefont
  {M.}~\bibnamefont {Baguelin}}, \bibinfo {author} {\bibfnamefont
  {S.}~\bibnamefont {Bhatia}}, \bibinfo {author} {\bibfnamefont
  {A.}~\bibnamefont {Boonyasiri}}, \bibinfo {author} {\bibfnamefont
  {Z.}~\bibnamefont {Cucunub{\'a}}}, \bibinfo {author} {\bibfnamefont
  {G.}~\bibnamefont {Cuomo-Dannenburg}}, \emph {et~al.},\ }\bibfield  {title}
  {\bibinfo {title} {Impact of non-pharmaceutical interventions ({NPI}s) to
  reduce {COVID-19} mortality and healthcare demand. 2020},\ }\href@noop {}
  {\bibfield  {journal} {\bibinfo  {journal} {DOI}\ }\textbf {\bibinfo {volume}
  {10}},\ \bibinfo {pages} {77482} (\bibinfo {year} {2020})}\BibitemShut
  {NoStop}%
\bibitem [{\citenamefont {Kraemer}\ \emph {et~al.}(2020)\citenamefont
  {Kraemer}, \citenamefont {Yang}, \citenamefont {Gutierrez}, \citenamefont
  {Wu}, \citenamefont {Klein}, \citenamefont {Pigott}, \citenamefont
  {Du~Plessis}, \citenamefont {Faria}, \citenamefont {Li}, \citenamefont
  {Hanage} \emph {et~al.}}]{kraemer2020effect}%
  \BibitemOpen
  \bibfield  {author} {\bibinfo {author} {\bibfnamefont {M.~U.}\ \bibnamefont
  {Kraemer}}, \bibinfo {author} {\bibfnamefont {C.-H.}\ \bibnamefont {Yang}},
  \bibinfo {author} {\bibfnamefont {B.}~\bibnamefont {Gutierrez}}, \bibinfo
  {author} {\bibfnamefont {C.-H.}\ \bibnamefont {Wu}}, \bibinfo {author}
  {\bibfnamefont {B.}~\bibnamefont {Klein}}, \bibinfo {author} {\bibfnamefont
  {D.~M.}\ \bibnamefont {Pigott}}, \bibinfo {author} {\bibfnamefont
  {L.}~\bibnamefont {Du~Plessis}}, \bibinfo {author} {\bibfnamefont {N.~R.}\
  \bibnamefont {Faria}}, \bibinfo {author} {\bibfnamefont {R.}~\bibnamefont
  {Li}}, \bibinfo {author} {\bibfnamefont {W.~P.}\ \bibnamefont {Hanage}},
  \emph {et~al.},\ }\bibfield  {title} {\bibinfo {title} {The effect of human
  mobility and control measures on the {COVID-19} epidemic in {C}hina},\
  }\href@noop {} {\bibfield  {journal} {\bibinfo  {journal} {Science}\ }\textbf
  {\bibinfo {volume} {368}},\ \bibinfo {pages} {493} (\bibinfo {year}
  {2020})}\BibitemShut {NoStop}%
\bibitem [{\citenamefont {Wang}\ \emph {et~al.}(2020)\citenamefont {Wang},
  \citenamefont {Ng},\ and\ \citenamefont {Brook}}]{wang2020response}%
  \BibitemOpen
  \bibfield  {author} {\bibinfo {author} {\bibfnamefont {C.~J.}\ \bibnamefont
  {Wang}}, \bibinfo {author} {\bibfnamefont {C.~Y.}\ \bibnamefont {Ng}},\ and\
  \bibinfo {author} {\bibfnamefont {R.~H.}\ \bibnamefont {Brook}},\ }\bibfield
  {title} {\bibinfo {title} {Response to {COVID-19} in {T}aiwan: big data
  analytics, new technology, and proactive testing},\ }\href@noop {} {\bibfield
   {journal} {\bibinfo  {journal} {Jama}\ }\textbf {\bibinfo {volume} {323}},\
  \bibinfo {pages} {1341} (\bibinfo {year} {2020})}\BibitemShut {NoStop}%
\bibitem [{\citenamefont {Kucharski}\ \emph {et~al.}(2020)\citenamefont
  {Kucharski}, \citenamefont {Russell}, \citenamefont {Diamond}, \citenamefont
  {Liu}, \citenamefont {Edmunds}, \citenamefont {Funk}, \citenamefont {Eggo},
  \citenamefont {Sun}, \citenamefont {Jit}, \citenamefont {Munday} \emph
  {et~al.}}]{kucharski2020early}%
  \BibitemOpen
  \bibfield  {author} {\bibinfo {author} {\bibfnamefont {A.~J.}\ \bibnamefont
  {Kucharski}}, \bibinfo {author} {\bibfnamefont {T.~W.}\ \bibnamefont
  {Russell}}, \bibinfo {author} {\bibfnamefont {C.}~\bibnamefont {Diamond}},
  \bibinfo {author} {\bibfnamefont {Y.}~\bibnamefont {Liu}}, \bibinfo {author}
  {\bibfnamefont {J.}~\bibnamefont {Edmunds}}, \bibinfo {author} {\bibfnamefont
  {S.}~\bibnamefont {Funk}}, \bibinfo {author} {\bibfnamefont {R.~M.}\
  \bibnamefont {Eggo}}, \bibinfo {author} {\bibfnamefont {F.}~\bibnamefont
  {Sun}}, \bibinfo {author} {\bibfnamefont {M.}~\bibnamefont {Jit}}, \bibinfo
  {author} {\bibfnamefont {J.~D.}\ \bibnamefont {Munday}}, \emph {et~al.},\
  }\bibfield  {title} {\bibinfo {title} {Early dynamics of transmission and
  control of {COVID-19}: a mathematical modelling study},\ }\href@noop {}
  {\bibfield  {journal} {\bibinfo  {journal} {The Lancet Infectious Diseases}\
  } (\bibinfo {year} {2020})}\BibitemShut {NoStop}%
\bibitem [{\citenamefont {Overton}\ \emph {et~al.}(2020)\citenamefont
  {Overton}, \citenamefont {Stage}, \citenamefont {Ahmad}, \citenamefont
  {Curran-Sebastian}, \citenamefont {Dark}, \citenamefont {Das}, \citenamefont
  {Fearon}, \citenamefont {Felton}, \citenamefont {Fyles}, \citenamefont {Gent}
  \emph {et~al.}}]{overton2020using}%
  \BibitemOpen
  \bibfield  {author} {\bibinfo {author} {\bibfnamefont {C.~E.}\ \bibnamefont
  {Overton}}, \bibinfo {author} {\bibfnamefont {H.~B.}\ \bibnamefont {Stage}},
  \bibinfo {author} {\bibfnamefont {S.}~\bibnamefont {Ahmad}}, \bibinfo
  {author} {\bibfnamefont {J.}~\bibnamefont {Curran-Sebastian}}, \bibinfo
  {author} {\bibfnamefont {P.}~\bibnamefont {Dark}}, \bibinfo {author}
  {\bibfnamefont {R.}~\bibnamefont {Das}}, \bibinfo {author} {\bibfnamefont
  {E.}~\bibnamefont {Fearon}}, \bibinfo {author} {\bibfnamefont
  {T.}~\bibnamefont {Felton}}, \bibinfo {author} {\bibfnamefont
  {M.}~\bibnamefont {Fyles}}, \bibinfo {author} {\bibfnamefont
  {N.}~\bibnamefont {Gent}}, \emph {et~al.},\ }\bibfield  {title} {\bibinfo
  {title} {Using statistics and mathematical modelling to understand infectious
  disease outbreaks: {COVID-19} as an example},\ }\href@noop {} {\bibfield
  {journal} {\bibinfo  {journal} {arXiv preprint arXiv:2005.04937}\ } (\bibinfo
  {year} {2020})}\BibitemShut {NoStop}%
\bibitem [{\citenamefont {Giordano}\ \emph {et~al.}(2020)\citenamefont
  {Giordano}, \citenamefont {Blanchini}, \citenamefont {Bruno}, \citenamefont
  {Colaneri}, \citenamefont {Di~Filippo}, \citenamefont {Di~Matteo},\ and\
  \citenamefont {Colaneri}}]{giordano2020modelling}%
  \BibitemOpen
  \bibfield  {author} {\bibinfo {author} {\bibfnamefont {G.}~\bibnamefont
  {Giordano}}, \bibinfo {author} {\bibfnamefont {F.}~\bibnamefont {Blanchini}},
  \bibinfo {author} {\bibfnamefont {R.}~\bibnamefont {Bruno}}, \bibinfo
  {author} {\bibfnamefont {P.}~\bibnamefont {Colaneri}}, \bibinfo {author}
  {\bibfnamefont {A.}~\bibnamefont {Di~Filippo}}, \bibinfo {author}
  {\bibfnamefont {A.}~\bibnamefont {Di~Matteo}},\ and\ \bibinfo {author}
  {\bibfnamefont {M.}~\bibnamefont {Colaneri}},\ }\bibfield  {title} {\bibinfo
  {title} {Modelling the {COVID-19} epidemic and implementation of
  population-wide interventions in {I}taly},\ }\href@noop {} {\bibfield
  {journal} {\bibinfo  {journal} {Nature Medicine}\ ,\ \bibinfo {pages} {1}}
  (\bibinfo {year} {2020})}\BibitemShut {NoStop}%
\bibitem [{\citenamefont {Prem}\ \emph {et~al.}(2020)\citenamefont {Prem},
  \citenamefont {Liu}, \citenamefont {Russell}, \citenamefont {Kucharski},
  \citenamefont {Eggo}, \citenamefont {Davies}, \citenamefont {Flasche},
  \citenamefont {Clifford}, \citenamefont {Pearson}, \citenamefont {Munday}
  \emph {et~al.}}]{prem2020effect}%
  \BibitemOpen
  \bibfield  {author} {\bibinfo {author} {\bibfnamefont {K.}~\bibnamefont
  {Prem}}, \bibinfo {author} {\bibfnamefont {Y.}~\bibnamefont {Liu}}, \bibinfo
  {author} {\bibfnamefont {T.~W.}\ \bibnamefont {Russell}}, \bibinfo {author}
  {\bibfnamefont {A.~J.}\ \bibnamefont {Kucharski}}, \bibinfo {author}
  {\bibfnamefont {R.~M.}\ \bibnamefont {Eggo}}, \bibinfo {author}
  {\bibfnamefont {N.}~\bibnamefont {Davies}}, \bibinfo {author} {\bibfnamefont
  {S.}~\bibnamefont {Flasche}}, \bibinfo {author} {\bibfnamefont
  {S.}~\bibnamefont {Clifford}}, \bibinfo {author} {\bibfnamefont {C.~A.}\
  \bibnamefont {Pearson}}, \bibinfo {author} {\bibfnamefont {J.~D.}\
  \bibnamefont {Munday}}, \emph {et~al.},\ }\bibfield  {title} {\bibinfo
  {title} {The effect of control strategies to reduce social mixing on outcomes
  of the {COVID-19} epidemic in {W}uhan, {C}hina: a modelling study},\
  }\href@noop {} {\bibfield  {journal} {\bibinfo  {journal} {The Lancet Public
  Health}\ } (\bibinfo {year} {2020})}\BibitemShut {NoStop}%
\bibitem [{\citenamefont {Gaeta}(2020)}]{gaeta2020simple}%
  \BibitemOpen
  \bibfield  {author} {\bibinfo {author} {\bibfnamefont {G.}~\bibnamefont
  {Gaeta}},\ }\bibfield  {title} {\bibinfo {title} {A simple {SIR} model with a
  large set of asymptomatic infectives},\ }\href@noop {} {\bibfield  {journal}
  {\bibinfo  {journal} {arXiv preprint arXiv:2003.08720}\ } (\bibinfo {year}
  {2020})}\BibitemShut {NoStop}%
\bibitem [{\citenamefont {Scabini}\ \emph {et~al.}(2020)\citenamefont
  {Scabini}, \citenamefont {Ribas}, \citenamefont {Neiva}, \citenamefont
  {Junior}, \citenamefont {Farf{\'a}n},\ and\ \citenamefont
  {Bruno}}]{scabini2020social}%
  \BibitemOpen
  \bibfield  {author} {\bibinfo {author} {\bibfnamefont {L.~F.}\ \bibnamefont
  {Scabini}}, \bibinfo {author} {\bibfnamefont {L.~C.}\ \bibnamefont {Ribas}},
  \bibinfo {author} {\bibfnamefont {M.~B.}\ \bibnamefont {Neiva}}, \bibinfo
  {author} {\bibfnamefont {A.~G.}\ \bibnamefont {Junior}}, \bibinfo {author}
  {\bibfnamefont {A.~J.}\ \bibnamefont {Farf{\'a}n}},\ and\ \bibinfo {author}
  {\bibfnamefont {O.~M.}\ \bibnamefont {Bruno}},\ }\bibfield  {title} {\bibinfo
  {title} {Social interaction layers in complex networks for the dynamical
  epidemic modeling of {COVID-19} in {B}razil},\ }\href@noop {} {\bibfield
  {journal} {\bibinfo  {journal} {arXiv preprint arXiv:2005.08125}\ } (\bibinfo
  {year} {2020})}\BibitemShut {NoStop}%
\bibitem [{\citenamefont {Leung}\ \emph {et~al.}(2020)\citenamefont {Leung},
  \citenamefont {Wu}, \citenamefont {Liu},\ and\ \citenamefont
  {Leung}}]{leung2020first}%
  \BibitemOpen
  \bibfield  {author} {\bibinfo {author} {\bibfnamefont {K.}~\bibnamefont
  {Leung}}, \bibinfo {author} {\bibfnamefont {J.~T.}\ \bibnamefont {Wu}},
  \bibinfo {author} {\bibfnamefont {D.}~\bibnamefont {Liu}},\ and\ \bibinfo
  {author} {\bibfnamefont {G.~M.}\ \bibnamefont {Leung}},\ }\bibfield  {title}
  {\bibinfo {title} {First-wave {COVID-19} transmissibility and severity in
  {C}hina outside {H}ubei after control measures, and second-wave scenario
  planning: a modelling impact assessment},\ }\href@noop {} {\bibfield
  {journal} {\bibinfo  {journal} {The Lancet}\ } (\bibinfo {year}
  {2020})}\BibitemShut {NoStop}%
\bibitem [{\citenamefont {Candido}\ \emph {et~al.}(2020)\citenamefont
  {Candido}, \citenamefont {Claro}, \citenamefont {de~Jesus}, \citenamefont
  {Souza}, \citenamefont {Moreira}, \citenamefont {Dellicour}, \citenamefont
  {Mellan}, \citenamefont {du~Plessis}, \citenamefont {Pereira}, \citenamefont
  {Sales} \emph {et~al.}}]{candido2020evolution}%
  \BibitemOpen
  \bibfield  {author} {\bibinfo {author} {\bibfnamefont {D.~S.}\ \bibnamefont
  {Candido}}, \bibinfo {author} {\bibfnamefont {I.~M.}\ \bibnamefont {Claro}},
  \bibinfo {author} {\bibfnamefont {J.~G.}\ \bibnamefont {de~Jesus}}, \bibinfo
  {author} {\bibfnamefont {W.~M.}\ \bibnamefont {Souza}}, \bibinfo {author}
  {\bibfnamefont {F.~R.}\ \bibnamefont {Moreira}}, \bibinfo {author}
  {\bibfnamefont {S.}~\bibnamefont {Dellicour}}, \bibinfo {author}
  {\bibfnamefont {T.~A.}\ \bibnamefont {Mellan}}, \bibinfo {author}
  {\bibfnamefont {L.}~\bibnamefont {du~Plessis}}, \bibinfo {author}
  {\bibfnamefont {R.~H.}\ \bibnamefont {Pereira}}, \bibinfo {author}
  {\bibfnamefont {F.~C.}\ \bibnamefont {Sales}}, \emph {et~al.},\ }\bibfield
  {title} {\bibinfo {title} {Evolution and epidemic spread of sars-cov-2 in
  {B}razil},\ }\href@noop {} {\bibfield  {journal} {\bibinfo  {journal}
  {Science}\ } (\bibinfo {year} {2020})}\BibitemShut {NoStop}%
\bibitem [{\citenamefont {Fang}\ \emph {et~al.}(2020)\citenamefont {Fang},
  \citenamefont {Nie},\ and\ \citenamefont {Penny}}]{fang2020transmission}%
  \BibitemOpen
  \bibfield  {author} {\bibinfo {author} {\bibfnamefont {Y.}~\bibnamefont
  {Fang}}, \bibinfo {author} {\bibfnamefont {Y.}~\bibnamefont {Nie}},\ and\
  \bibinfo {author} {\bibfnamefont {M.}~\bibnamefont {Penny}},\ }\bibfield
  {title} {\bibinfo {title} {Transmission dynamics of the {COVID-19} outbreak
  and effectiveness of government interventions: A data-driven analysis},\
  }\href@noop {} {\bibfield  {journal} {\bibinfo  {journal} {Journal of medical
  virology}\ }\textbf {\bibinfo {volume} {92}},\ \bibinfo {pages} {645}
  (\bibinfo {year} {2020})}\BibitemShut {NoStop}%
\bibitem [{\citenamefont {Massad}\ \emph {et~al.}(2020)\citenamefont {Massad},
  \citenamefont {Amaku}, \citenamefont {Wilder-Smith}, \citenamefont {dos
  Santos}, \citenamefont {Struchiner},\ and\ \citenamefont
  {Coutinho}}]{massad2020two}%
  \BibitemOpen
  \bibfield  {author} {\bibinfo {author} {\bibfnamefont {E.}~\bibnamefont
  {Massad}}, \bibinfo {author} {\bibfnamefont {M.}~\bibnamefont {Amaku}},
  \bibinfo {author} {\bibfnamefont {A.}~\bibnamefont {Wilder-Smith}}, \bibinfo
  {author} {\bibfnamefont {P.~C.~C.}\ \bibnamefont {dos Santos}}, \bibinfo
  {author} {\bibfnamefont {C.~J.}\ \bibnamefont {Struchiner}},\ and\ \bibinfo
  {author} {\bibfnamefont {F.~A.~B.}\ \bibnamefont {Coutinho}},\ }\bibfield
  {title} {\bibinfo {title} {Two complementary model-based methods for
  calculating the risk of international spreading of anovel virus from the
  outbreak epicentre. the case of {COVID-19}},\ }\href@noop {} {\bibfield
  {journal} {\bibinfo  {journal} {Epidemiology \& Infection}\ ,\ \bibinfo
  {pages} {1}} (\bibinfo {year} {2020})}\BibitemShut {NoStop}%
\bibitem [{\citenamefont {Sch{\"u}ttler}\ \emph {et~al.}(2020)\citenamefont
  {Sch{\"u}ttler}, \citenamefont {Schlickeiser}, \citenamefont {Schlickeiser},\
  and\ \citenamefont {Kr{\"o}ger}}]{schuttler2020covid}%
  \BibitemOpen
  \bibfield  {author} {\bibinfo {author} {\bibfnamefont {J.}~\bibnamefont
  {Sch{\"u}ttler}}, \bibinfo {author} {\bibfnamefont {R.}~\bibnamefont
  {Schlickeiser}}, \bibinfo {author} {\bibfnamefont {F.}~\bibnamefont
  {Schlickeiser}},\ and\ \bibinfo {author} {\bibfnamefont {M.}~\bibnamefont
  {Kr{\"o}ger}},\ }\bibfield  {title} {\bibinfo {title} {Covid-19 predictions
  using a {G}auss model, based on data from april 2},\ }\href@noop {}
  {\bibfield  {journal} {\bibinfo  {journal} {Physics}\ }\textbf {\bibinfo
  {volume} {2}},\ \bibinfo {pages} {197} (\bibinfo {year} {2020})}\BibitemShut
  {NoStop}%
\bibitem [{\citenamefont {Pedrosa}(2020)}]{pedrosa2020dynamics}%
  \BibitemOpen
  \bibfield  {author} {\bibinfo {author} {\bibfnamefont {R.~H.}\ \bibnamefont
  {Pedrosa}},\ }\bibfield  {title} {\bibinfo {title} {The dynamics of covid-19:
  weather, demographics and infection timeline},\ }\href@noop {} {\bibfield
  {journal} {\bibinfo  {journal} {medRxiv}\ } (\bibinfo {year}
  {2020})}\BibitemShut {NoStop}%
\bibitem [{\citenamefont {ul~Hassan}(2020)}]{ul2020understanding}%
  \BibitemOpen
  \bibfield  {author} {\bibinfo {author} {\bibfnamefont {A.~I.}\ \bibnamefont
  {ul~Hassan}},\ }\bibfield  {title} {\bibinfo {title} {Understanding the
  {COVID-19} pandemic curve through statistical approach},\ }\href@noop {}
  {\bibfield  {journal} {\bibinfo  {journal} {https://www. medrxiv.
  org/content/medrxiv/early/2020/04/08/2020.04. 06.20055426. full. pdf}\ }
  (\bibinfo {year} {2020})}\BibitemShut {NoStop}%
\bibitem [{\citenamefont {Xu}\ \emph {et~al.}(2020)\citenamefont {Xu},
  \citenamefont {Wang}, \citenamefont {Dong}, \citenamefont {Shen},\ and\
  \citenamefont {Xu}}]{xu2020analysis}%
  \BibitemOpen
  \bibfield  {author} {\bibinfo {author} {\bibfnamefont {X.}~\bibnamefont
  {Xu}}, \bibinfo {author} {\bibfnamefont {S.}~\bibnamefont {Wang}}, \bibinfo
  {author} {\bibfnamefont {J.}~\bibnamefont {Dong}}, \bibinfo {author}
  {\bibfnamefont {Z.}~\bibnamefont {Shen}},\ and\ \bibinfo {author}
  {\bibfnamefont {S.}~\bibnamefont {Xu}},\ }\bibfield  {title} {\bibinfo
  {title} {An analysis of the domestic resumption of social production and life
  under the {COVID-19} epidemic},\ }\href@noop {} {\bibfield  {journal}
  {\bibinfo  {journal} {PloS one}\ }\textbf {\bibinfo {volume} {15}},\ \bibinfo
  {pages} {e0236387} (\bibinfo {year} {2020})}\BibitemShut {NoStop}%
\bibitem [{\citenamefont {Paix{\~a}o}\ \emph {et~al.}(2020)\citenamefont
  {Paix{\~a}o}, \citenamefont {Baroni}, \citenamefont {Salles}, \citenamefont
  {Escobar}, \citenamefont {de~Sousa}, \citenamefont {Pedroso}, \citenamefont
  {Saldanha}, \citenamefont {Coutinho}, \citenamefont {Porto},\ and\
  \citenamefont {Ogasawara}}]{paixao2020estimation}%
  \BibitemOpen
  \bibfield  {author} {\bibinfo {author} {\bibfnamefont {B.}~\bibnamefont
  {Paix{\~a}o}}, \bibinfo {author} {\bibfnamefont {L.}~\bibnamefont {Baroni}},
  \bibinfo {author} {\bibfnamefont {R.}~\bibnamefont {Salles}}, \bibinfo
  {author} {\bibfnamefont {L.}~\bibnamefont {Escobar}}, \bibinfo {author}
  {\bibfnamefont {C.}~\bibnamefont {de~Sousa}}, \bibinfo {author}
  {\bibfnamefont {M.}~\bibnamefont {Pedroso}}, \bibinfo {author} {\bibfnamefont
  {R.}~\bibnamefont {Saldanha}}, \bibinfo {author} {\bibfnamefont
  {R.}~\bibnamefont {Coutinho}}, \bibinfo {author} {\bibfnamefont
  {F.}~\bibnamefont {Porto}},\ and\ \bibinfo {author} {\bibfnamefont
  {E.}~\bibnamefont {Ogasawara}},\ }\bibfield  {title} {\bibinfo {title}
  {Estimation of {COVID-19} under-reporting in {B}razilian {S}tates through
  sari},\ }\href@noop {} {\bibfield  {journal} {\bibinfo  {journal} {arXiv
  preprint arXiv:2006.12759}\ } (\bibinfo {year} {2020})}\BibitemShut {NoStop}%
\bibitem [{\citenamefont {Krantz}\ and\ \citenamefont
  {Rao}(2020)}]{krantz2020level}%
  \BibitemOpen
  \bibfield  {author} {\bibinfo {author} {\bibfnamefont {S.~G.}\ \bibnamefont
  {Krantz}}\ and\ \bibinfo {author} {\bibfnamefont {A.~S.~S.}\ \bibnamefont
  {Rao}},\ }\bibfield  {title} {\bibinfo {title} {Level of underreporting
  including underdiagnosis before the first peak of {COVID-19} in various
  countries: Preliminary retrospective results based on wavelets and
  deterministic modeling},\ }\href@noop {} {\bibfield  {journal} {\bibinfo
  {journal} {Infection Control \& Hospital Epidemiology}\ ,\ \bibinfo {pages}
  {1}} (\bibinfo {year} {2020})}\BibitemShut {NoStop}%
\bibitem [{\citenamefont {Jagodnik}\ \emph {et~al.}(2020)\citenamefont
  {Jagodnik}, \citenamefont {Ray}, \citenamefont {Giorgi},\ and\ \citenamefont
  {Lachmann}}]{jagodnik2020correcting}%
  \BibitemOpen
  \bibfield  {author} {\bibinfo {author} {\bibfnamefont {K.}~\bibnamefont
  {Jagodnik}}, \bibinfo {author} {\bibfnamefont {F.}~\bibnamefont {Ray}},
  \bibinfo {author} {\bibfnamefont {F.~M.}\ \bibnamefont {Giorgi}},\ and\
  \bibinfo {author} {\bibfnamefont {A.}~\bibnamefont {Lachmann}},\ }\bibfield
  {title} {\bibinfo {title} {Correcting under-reported {COVID-19} case numbers:
  estimating the true scale of the pandemic},\ }\href@noop {} {\bibfield
  {journal} {\bibinfo  {journal} {Preprint medRvix}\ }\textbf {\bibinfo
  {volume} {14}} (\bibinfo {year} {2020})}\BibitemShut {NoStop}%
\bibitem [{\citenamefont {Cohen}\ and\ \citenamefont
  {Kupferschmidt}(2020)}]{cohen2020countries}%
  \BibitemOpen
  \bibfield  {author} {\bibinfo {author} {\bibfnamefont {J.}~\bibnamefont
  {Cohen}}\ and\ \bibinfo {author} {\bibfnamefont {K.}~\bibnamefont
  {Kupferschmidt}},\ }\href@noop {} {\bibinfo {title} {Countries test tactics
  in ‘war’against {COVID-19}}} (\bibinfo {year} {2020})\BibitemShut
  {NoStop}%
\bibitem [{\citenamefont {Capaldi}\ \emph {et~al.}(2012)\citenamefont
  {Capaldi}, \citenamefont {Behrend}, \citenamefont {Berman}, \citenamefont
  {Smith}, \citenamefont {Wright},\ and\ \citenamefont
  {Lloyd}}]{capaldi2012parameter}%
  \BibitemOpen
  \bibfield  {author} {\bibinfo {author} {\bibfnamefont {A.}~\bibnamefont
  {Capaldi}}, \bibinfo {author} {\bibfnamefont {S.}~\bibnamefont {Behrend}},
  \bibinfo {author} {\bibfnamefont {B.}~\bibnamefont {Berman}}, \bibinfo
  {author} {\bibfnamefont {J.}~\bibnamefont {Smith}}, \bibinfo {author}
  {\bibfnamefont {J.}~\bibnamefont {Wright}},\ and\ \bibinfo {author}
  {\bibfnamefont {A.~L.}\ \bibnamefont {Lloyd}},\ }\bibfield  {title} {\bibinfo
  {title} {Parameter estimation and uncertainty quantication for an epidemic
  model},\ }\href@noop {} {\bibfield  {journal} {\bibinfo  {journal}
  {Mathematical biosciences and engineering}\ ,\ \bibinfo {pages} {553}}
  (\bibinfo {year} {2012})}\BibitemShut {NoStop}%
\bibitem [{\citenamefont {Gaffey}\ and\ \citenamefont
  {Viboud}(2018)}]{gaffey2018application}%
  \BibitemOpen
  \bibfield  {author} {\bibinfo {author} {\bibfnamefont {R.~H.}\ \bibnamefont
  {Gaffey}}\ and\ \bibinfo {author} {\bibfnamefont {C.}~\bibnamefont
  {Viboud}},\ }\bibfield  {title} {\bibinfo {title} {Application of the {CDC}
  {E}bola {R}esponse modeling tool to disease predictions},\ }\href@noop {}
  {\bibfield  {journal} {\bibinfo  {journal} {Epidemics}\ }\textbf {\bibinfo
  {volume} {22}},\ \bibinfo {pages} {22} (\bibinfo {year} {2018})}\BibitemShut
  {NoStop}%
\bibitem [{\citenamefont {Kuniya}(2020)}]{kuniya2020prediction}%
  \BibitemOpen
  \bibfield  {author} {\bibinfo {author} {\bibfnamefont {T.}~\bibnamefont
  {Kuniya}},\ }\bibfield  {title} {\bibinfo {title} {Prediction of the epidemic
  peak of coronavirus disease in japan, 2020},\ }\href@noop {} {\bibfield
  {journal} {\bibinfo  {journal} {Journal of clinical medicine}\ }\textbf
  {\bibinfo {volume} {9}},\ \bibinfo {pages} {789} (\bibinfo {year}
  {2020})}\BibitemShut {NoStop}%
\bibitem [{\citenamefont {Smirnova}\ \emph {et~al.}(2019)\citenamefont
  {Smirnova}, \citenamefont {Sirb},\ and\ \citenamefont
  {Chowell}}]{smirnova2019stable}%
  \BibitemOpen
  \bibfield  {author} {\bibinfo {author} {\bibfnamefont {A.}~\bibnamefont
  {Smirnova}}, \bibinfo {author} {\bibfnamefont {B.}~\bibnamefont {Sirb}},\
  and\ \bibinfo {author} {\bibfnamefont {G.}~\bibnamefont {Chowell}},\
  }\bibfield  {title} {\bibinfo {title} {On stable parameter estimation and
  forecasting in epidemiology by the levenberg--marquardt algorithm with
  broyden’s rank-one updates for the jacobian operator},\ }\href@noop {}
  {\bibfield  {journal} {\bibinfo  {journal} {Bulletin of mathematical
  biology}\ }\textbf {\bibinfo {volume} {81}},\ \bibinfo {pages} {4210}
  (\bibinfo {year} {2019})}\BibitemShut {NoStop}%
\bibitem [{\citenamefont {Peto}(2020)}]{peto2020covid}%
  \BibitemOpen
  \bibfield  {author} {\bibinfo {author} {\bibfnamefont {J.}~\bibnamefont
  {Peto}},\ }\bibfield  {title} {\bibinfo {title} {Covid-19 mass testing
  facilities could end the epidemic rapidly},\ }\href@noop {} {\bibfield
  {journal} {\bibinfo  {journal} {Bmj}\ }\textbf {\bibinfo {volume} {368}}
  (\bibinfo {year} {2020})}\BibitemShut {NoStop}%
\bibitem [{\citenamefont {Flaxman}\ \emph {et~al.}(2020)\citenamefont
  {Flaxman}, \citenamefont {Mishra}, \citenamefont {Gandy}, \citenamefont
  {Unwin}, \citenamefont {Coupland}, \citenamefont {Mellan}, \citenamefont
  {Zhu}, \citenamefont {Berah}, \citenamefont {Eaton}, \citenamefont
  {Perez~Guzman} \emph {et~al.}}]{flaxman2020report}%
  \BibitemOpen
  \bibfield  {author} {\bibinfo {author} {\bibfnamefont {S.}~\bibnamefont
  {Flaxman}}, \bibinfo {author} {\bibfnamefont {S.}~\bibnamefont {Mishra}},
  \bibinfo {author} {\bibfnamefont {A.}~\bibnamefont {Gandy}}, \bibinfo
  {author} {\bibfnamefont {H.}~\bibnamefont {Unwin}}, \bibinfo {author}
  {\bibfnamefont {H.}~\bibnamefont {Coupland}}, \bibinfo {author}
  {\bibfnamefont {T.}~\bibnamefont {Mellan}}, \bibinfo {author} {\bibfnamefont
  {H.}~\bibnamefont {Zhu}}, \bibinfo {author} {\bibfnamefont {T.}~\bibnamefont
  {Berah}}, \bibinfo {author} {\bibfnamefont {J.}~\bibnamefont {Eaton}},
  \bibinfo {author} {\bibfnamefont {P.}~\bibnamefont {Perez~Guzman}}, \emph
  {et~al.},\ }\href@noop {} {\emph {\bibinfo {title} {Report 13: Estimating the
  number of infections and the impact of non-pharmaceutical interventions on
  {COVID-19} in 11 European countries}}},\ \bibinfo {type} {Tech. Rep.}\
  (\bibinfo {year} {2020})\BibitemShut {NoStop}%
\bibitem [{\citenamefont {Verity}\ \emph {et~al.}(2020)\citenamefont {Verity},
  \citenamefont {Okell}, \citenamefont {Dorigatti}, \citenamefont {Winskill},
  \citenamefont {Whittaker}, \citenamefont {Imai}, \citenamefont
  {Cuomo-Dannenburg}, \citenamefont {Thompson}, \citenamefont {Walker},
  \citenamefont {Fu} \emph {et~al.}}]{verity2020estimates}%
  \BibitemOpen
  \bibfield  {author} {\bibinfo {author} {\bibfnamefont {R.}~\bibnamefont
  {Verity}}, \bibinfo {author} {\bibfnamefont {L.~C.}\ \bibnamefont {Okell}},
  \bibinfo {author} {\bibfnamefont {I.}~\bibnamefont {Dorigatti}}, \bibinfo
  {author} {\bibfnamefont {P.}~\bibnamefont {Winskill}}, \bibinfo {author}
  {\bibfnamefont {C.}~\bibnamefont {Whittaker}}, \bibinfo {author}
  {\bibfnamefont {N.}~\bibnamefont {Imai}}, \bibinfo {author} {\bibfnamefont
  {G.}~\bibnamefont {Cuomo-Dannenburg}}, \bibinfo {author} {\bibfnamefont
  {H.}~\bibnamefont {Thompson}}, \bibinfo {author} {\bibfnamefont {P.~G.}\
  \bibnamefont {Walker}}, \bibinfo {author} {\bibfnamefont {H.}~\bibnamefont
  {Fu}}, \emph {et~al.},\ }\bibfield  {title} {\bibinfo {title} {Estimates of
  the severity of coronavirus disease 2019: a model-based analysis},\
  }\href@noop {} {\bibfield  {journal} {\bibinfo  {journal} {The Lancet
  infectious diseases}\ } (\bibinfo {year} {2020})}\BibitemShut {NoStop}%
\bibitem [{Note1()}]{Note1}%
  \BibitemOpen
  \bibinfo {note} {The average elapsed time from symptom onset to death was
  corrected to 17.8 days in \protect \cite {verity2020estimates}, instead of
  their previous estimates of 18.8 days used by \protect \cite
  {flaxman2020report}}\BibitemShut {NoStop}%
\bibitem [{Note2()}]{Note2}%
  \BibitemOpen
  \bibinfo {note} {The deaths data included in the analysis were taken from the
  European Centre for Disease Prevention and Control webpage:
  https://www.ecdc.europa.eu/en/publications-data/download-todays-data-geographic-distribution-covid-19-cases-worldwide}\BibitemShut
  {NoStop}%
\bibitem [{\citenamefont {Vanin}\ and\ \citenamefont
  {Helene}(2007)}]{vanin2007covariance}%
  \BibitemOpen
  \bibfield  {author} {\bibinfo {author} {\bibfnamefont {V.}~\bibnamefont
  {Vanin}}\ and\ \bibinfo {author} {\bibfnamefont {O.}~\bibnamefont {Helene}},\
  }\bibfield  {title} {\bibinfo {title} {Covariance analysis by means of the
  least squares method},\ }in\ \href@noop {} {\emph {\bibinfo {booktitle}
  {Update of X ray and gamma ray decay data standards for detector calibration
  and other applications. V. 2: Data selection, assessment and evaluation
  procedures}}}\ (\bibinfo {year} {2007})\BibitemShut {NoStop}%
\bibitem [{\citenamefont {Helene}\ \emph {et~al.}(2016)\citenamefont {Helene},
  \citenamefont {Mariano},\ and\ \citenamefont
  {Guimaraes-Filho}}]{helene2016useful}%
  \BibitemOpen
  \bibfield  {author} {\bibinfo {author} {\bibfnamefont {O.}~\bibnamefont
  {Helene}}, \bibinfo {author} {\bibfnamefont {L.}~\bibnamefont {Mariano}},\
  and\ \bibinfo {author} {\bibfnamefont {Z.}~\bibnamefont {Guimaraes-Filho}},\
  }\bibfield  {title} {\bibinfo {title} {Useful and little-known applications
  of the least square method and some consequences of covariances},\
  }\href@noop {} {\bibfield  {journal} {\bibinfo  {journal} {Nuclear
  Instruments and Methods in Physics Research Section A: Accelerators,
  Spectrometers, Detectors and Associated Equipment}\ }\textbf {\bibinfo
  {volume} {833}},\ \bibinfo {pages} {82} (\bibinfo {year} {2016})}\BibitemShut
  {NoStop}%
\bibitem [{\citenamefont {Bendavid}\ \emph {et~al.}(2020)\citenamefont
  {Bendavid}, \citenamefont {Mulaney}, \citenamefont {Sood}, \citenamefont
  {Shah}, \citenamefont {Ling}, \citenamefont {Bromley-Dulfano}, \citenamefont
  {Lai}, \citenamefont {Weissberg}, \citenamefont {Saavedra}, \citenamefont
  {Tedrow} \emph {et~al.}}]{bendavid2020covid}%
  \BibitemOpen
  \bibfield  {author} {\bibinfo {author} {\bibfnamefont {E.}~\bibnamefont
  {Bendavid}}, \bibinfo {author} {\bibfnamefont {B.}~\bibnamefont {Mulaney}},
  \bibinfo {author} {\bibfnamefont {N.}~\bibnamefont {Sood}}, \bibinfo {author}
  {\bibfnamefont {S.}~\bibnamefont {Shah}}, \bibinfo {author} {\bibfnamefont
  {E.}~\bibnamefont {Ling}}, \bibinfo {author} {\bibfnamefont {R.}~\bibnamefont
  {Bromley-Dulfano}}, \bibinfo {author} {\bibfnamefont {C.}~\bibnamefont
  {Lai}}, \bibinfo {author} {\bibfnamefont {Z.}~\bibnamefont {Weissberg}},
  \bibinfo {author} {\bibfnamefont {R.}~\bibnamefont {Saavedra}}, \bibinfo
  {author} {\bibfnamefont {J.}~\bibnamefont {Tedrow}}, \emph {et~al.},\
  }\bibfield  {title} {\bibinfo {title} {{COVID-19} antibody seroprevalence in
  {S}anta {C}lara {C}ounty, {C}alifornia},\ }\href@noop {} {\bibfield
  {journal} {\bibinfo  {journal} {MedRxiv}\ } (\bibinfo {year}
  {2020})}\BibitemShut {NoStop}%
\bibitem [{\citenamefont {Hallal}\ \emph {et~al.}(2020)\citenamefont {Hallal},
  \citenamefont {Hartwig}, \citenamefont {Horta}, \citenamefont {Victora},
  \citenamefont {Silveira}, \citenamefont {Struchiner}, \citenamefont
  {Vidaletti}, \citenamefont {Neumann}, \citenamefont {Pellanda}, \citenamefont
  {Dellagostin} \emph {et~al.}}]{hallal2020remarkable}%
  \BibitemOpen
  \bibfield  {author} {\bibinfo {author} {\bibfnamefont {P.}~\bibnamefont
  {Hallal}}, \bibinfo {author} {\bibfnamefont {F.}~\bibnamefont {Hartwig}},
  \bibinfo {author} {\bibfnamefont {B.}~\bibnamefont {Horta}}, \bibinfo
  {author} {\bibfnamefont {G.~D.}\ \bibnamefont {Victora}}, \bibinfo {author}
  {\bibfnamefont {M.}~\bibnamefont {Silveira}}, \bibinfo {author}
  {\bibfnamefont {C.}~\bibnamefont {Struchiner}}, \bibinfo {author}
  {\bibfnamefont {L.~P.}\ \bibnamefont {Vidaletti}}, \bibinfo {author}
  {\bibfnamefont {N.}~\bibnamefont {Neumann}}, \bibinfo {author} {\bibfnamefont
  {L.~C.}\ \bibnamefont {Pellanda}}, \bibinfo {author} {\bibfnamefont {O.~A.}\
  \bibnamefont {Dellagostin}}, \emph {et~al.},\ }\bibfield  {title} {\bibinfo
  {title} {Remarkable variability in sars-cov-2 antibodies across {B}razilian
  regions: nationwide serological household survey in 27 {s}tates},\
  }\href@noop {} {\bibfield  {journal} {\bibinfo  {journal} {medRxiv}\ }
  (\bibinfo {year} {2020})}\BibitemShut {NoStop}%
\bibitem [{\citenamefont {Gomes}\ \emph {et~al.}(2020)\citenamefont {Gomes},
  \citenamefont {Cerutti}, \citenamefont {Zandonade}, \citenamefont {Maciel},
  \citenamefont {de~Alencar}, \citenamefont {Almada}, \citenamefont {Cardoso},
  \citenamefont {Jabor}, \citenamefont {Zanotti}, \citenamefont {Reuter} \emph
  {et~al.}}]{gomes2020population}%
  \BibitemOpen
  \bibfield  {author} {\bibinfo {author} {\bibfnamefont {C.~C.}\ \bibnamefont
  {Gomes}}, \bibinfo {author} {\bibfnamefont {C.}~\bibnamefont {Cerutti}},
  \bibinfo {author} {\bibfnamefont {E.}~\bibnamefont {Zandonade}}, \bibinfo
  {author} {\bibfnamefont {E.~L.~N.}\ \bibnamefont {Maciel}}, \bibinfo {author}
  {\bibfnamefont {F.~E.~C.}\ \bibnamefont {de~Alencar}}, \bibinfo {author}
  {\bibfnamefont {G.~L.}\ \bibnamefont {Almada}}, \bibinfo {author}
  {\bibfnamefont {O.~A.}\ \bibnamefont {Cardoso}}, \bibinfo {author}
  {\bibfnamefont {P.~M.}\ \bibnamefont {Jabor}}, \bibinfo {author}
  {\bibfnamefont {R.~L.}\ \bibnamefont {Zanotti}}, \bibinfo {author}
  {\bibfnamefont {T.~Q.}\ \bibnamefont {Reuter}}, \emph {et~al.},\ }\bibfield
  {title} {\bibinfo {title} {A population-based study of the prevalence of
  {COVID-19} infection in {E}spirito {S}anto, {B}razil: methodology and results
  of the first stage},\ }\href@noop {} {\bibfield  {journal} {\bibinfo
  {journal} {medRxiv}\ } (\bibinfo {year} {2020})}\BibitemShut {NoStop}%
\end{thebibliography}%

\begin{table*}[b]
\caption{\label{tab:table1}
Best fit parameters of the Gompertz functions of the reconstructed infection curves and fitting results of the death's data obtained for all six countries. Also shown the 95\% confidence interval (CI) for the total number of deaths for Aug 31, 2020 and the estimated prevalences at the first day of each month.}
\begin{ruledtabular}
\begin{tabular}{ccccccc}
 &USA & Brazil & Mexico
 &UK & India\footnotemark[1] & Russia\footnotemark[1] \\
\hline
$N_1(10^{6})$& 11.6$\pm$1.1 & 12$\pm$14 & 11.9$\pm$0.7 &4.9$\pm$1.2 & 0.3$\pm$0.4 & 2.81$\pm$0.14\\
$\lambda _{1}(d^{-1})$& 0.109$\pm$0.012 & 0.030$\pm$0.009 & 0.0220$\pm$0.0010 & 0.100$\pm$0.017 & 0.06$\pm$0.03 & 0.0260$\pm$0.0011\\
$t_{01}(d)$& 65.7$\pm$1.0 & 76$\pm$25 & 103$\pm$3 & 51.8$\pm$2.2 & 68$\pm$13 & 112.8$\pm$2.2\\
$N_2(10^{6})$& 12.8$\pm$1.2 & 11$\pm$23 & - &2.1$\pm$1.3 & 19$\pm$8 & -\\
$\lambda _{2}(d^{-1})$& 0.026$\pm$0.004 & 0.023$\pm$0.022 & - & 0.036$\pm$0.010 & 0.017$\pm$0.005 &-\\
$t_{02}(d)$& 115$\pm$11 & 135$\pm$28 &-&  93$\pm$24  & 180$\pm$21 & -\\
$N_3(10^{6})$& 6$\pm$8 & - & - & - &- & -\\
$\lambda _{3}(d^{-1})$&  0.07$\pm$0.07 & - & - & - & - & -\\
$t_{03}(d)$&179$\pm$18 & - & - & - & - & -\\
$N_{d}$ (95\% CI) $(10^{3})$\footnotemark[2]& 180(160-220) & 120(110-130) & 60(59-62) & 47.0(46.6-47.3) & 59(54-63) & 16.3(16.0-16.7)\\
Prevalences (\%) & &  &  &  & \\
Mar 1, 2020 &0.4(0.2-0.7) & $ <$ 0.1 &$ <$ 0.1& $ <$ 0.1 & $ <$ 0.1 & $ <$ 0.1\\
Apr 1, 2020 &2.3(1.9-2.7) &0.2(0.15-0.23)& $ <$ 0.1& 5.0(4.1-5.9) & $ <$ 0.1 & $ <$ 0.1\\
May 1, 2020 &4.4(3.7-5.1) &1.4(1.3-1.6)& 0.8(0.7-0.9) & 8.3(6.7-9.9) & $ <$ 0.1 & 0.3(0.3-0.4)\\
Jun 1, 2020 &5.6(4.5-6.5) &3.7(3.3-4.1) & 2.7(2.4-2.9) & 9.6(7.8-11.5) & 0.1(0.1-0.2) & 0.9(0.8-0.9)\\
Jul 1, 2020 &6.6(5.4-7.8) &6.0(5.3-6.7) & 4.9(4.5-5.3) &10.3(8.3-12.2) & 0.3(0.3-0.4) & 1.3(1.2-1.5)\\
Aug 1, 2020 &8.3(5.7-10.9) &7.9(6.6-9.2) & 6.7(6.0-7.5) &10.5(8.4-12.6) & 0.6(0.5-0.7) & 1.6(1.5-1.8)\\
\textit{n/n.d.f.}\footnotemark[3]& 22/13 & 20/14 & 19/16 & 21/15 & 19/13 & 19/16\\
$ \chi ^{2}/p $ & 10.3/0.670 & 11.8/0.625 & 12.3/0.722 & 9.6/0.846 & 17.7/0.171 & 23.5/0.101\\
Begin date\footnotemark[4] & Mar 07, 2020 & Mar 21, 2020 & Mar 28, 2020 & Mar 14, 2020 &Mar 28, 2020 & Mar 28, 2020\\

\end{tabular}
\end{ruledtabular}
\footnotetext[1]{5\% error was added to the deaths data to achieve a successful fitting}
\footnotetext[2]{Model estimates for Aug 31, 2020}
\footnotetext[3]{\textit{n.d.f.}: number of degrees of freedom}
\footnotetext[4]{These dates correspond to the first day of the first epidemiological week included in the fitting.}
\end{table*}

\end{document}